\documentstyle[12pt]{article}
\begin{document}

\date{}

\title{Time-reversal-odd asymmetry in pion leptoproduction at HERMES}

\author{K.A. Oganessyan\thanks{On leave of absence from Yerevan Physics 
Institute, Alikhanian Br.2, 375036 Yerevan, Armenia} 
\thanks{to appear in the proceedings of DIS'98, Brussels, 4-8 April'98} \\
{\normalsize LNF-INFN, C.P. 13, Enrico Fermi 40, Frascati, Italy} 
}
\maketitle

\bigskip

\begin{abstract}
We estimate the size of the azimuthal asymmetry of 
the semi-inclusive pion production in the deep-inelastic scattering of a 
polarized lepton beam off an unpolarized nucleon target for the produced 
hadron, arising due to both nonperturbative and perturbative effects.
\end{abstract}

We analyse numerically the dependence of the azimuthal asymmetry of single 
inclusive charged pion production in $\vec{e}p$-scattering, shows up as a 
$\langle \sin\phi \rangle$ (related to the time-reversal-odd structures) 
on the transverse momentum cutoff $P_C$ at HERMES energies, with taking into 
account nonperturbative higher-twist \cite{LM} and perturbative 
$\alpha_S^2$-order QCD \cite{HAG} contributions. 

The leading non-zero contribution inducing a $\sin\phi$-dependence into the 
single hadron inclusive cross section arises from the absorptive part of 
the one-loop corrections to the $\gamma^{*}q \to qg$ and $\gamma^{*}g 
\to q\bar{q}$ subprocesses \cite{HAG}. Although 
that predicted $P_T$ integrated asymmetry,  $\langle \sin\phi \rangle$, 
is formally free from mass singularities, it may not be free from 
ambiguities due to the nonperturbative hadronic final-state interactions 
\cite{LM} which could be 
significant in small-$P_T$ regions. In this respect, introduction of a cutoff 
in outgoing hadron's transverse momentum $P_T$ \cite{CES} may be 
desirable not only from the experimental, but also from the theoretical point 
of view.  

The single asymmetry, $\langle \sin\phi \rangle$, in the outgoing hadron 
momentum distribution with respect to the lepton scattering plan is related 
to the left ($0 < \phi < \pi$)-right ($\pi < \phi < 2\pi$) asymmetry 
$$
A(Q^2, x_H, y, z_H, \vec{P}_T) = \frac{ 
{{d\sigma(left)} \over 
{dx_Hdydz_HdP^2_T}} - {{d\sigma(right)} \over {dx_Hdydz_HdP^2_T}}
}
{ 
{{d\sigma(left)} \over {dx_Hdydz_HdP^2_T}} + {{d\sigma(right)} 
\over {dx_Hdydz_HdP^2_T}}
},
$$ 
which is simply $4/{\pi}$  times $\langle \sin\phi \rangle$.  

Let us define the quantity $\langle \sin\phi \rangle$ as 
\begin{equation}
\label{R1}
\langle \sin\phi \rangle = \frac{\int d\sigma^{(HT)}\sin\phi + \int 
d\sigma^{(2)}\sin\phi }{\int d\sigma^{(0)} + \int d\sigma^{(1)}},
\end{equation}
where $d\sigma^{(HT)}$, $d\sigma^{(0)}$, $d\sigma^{(1)}$ and $d\sigma^{(2)}$ 
are the higher twist, lowest-order, first-order and second-order in $\alpha_S$
hadronic scattering cross sections expressed as a convolution of the partonic 
cross section and the distribution and fragmentation functions.

In order to estimate the expected asymmetry at HERMES defined by Eq(\ref{R1}), 
we have used the distribution functions \cite{GRV} and the hadronic 
fragmentation functions \cite{BKK} with Gaussian parametrization. For 
calculation of the higher twist 
contributions to the ``time-odd'' asymmetry \cite{LM}, we have used the 
twist-tree distribution and twist-two ``time-odd'' fragmentation 
function \cite{JMR} \footnote{in combination with Collins parameterization 
\cite{COL}}. Our numerical results for the left-handed electron at HERMES 
energies are shown in Fig.1. The asymmetry for the right-handed electron 
scattering is obtained by just reversing the sign in the figure. The 
measurement of such a single spin asymmetry can allow to determine the 
naively-time-odd quark fragmentation functions. It is important, however, 
to have good particle identification and sufficient azimuthal resolution 
in the forward direction. 

I wish to thank H. Avakian, N. Bianchi, A. Kotzinian and P. Mulders for 
useful discussions.

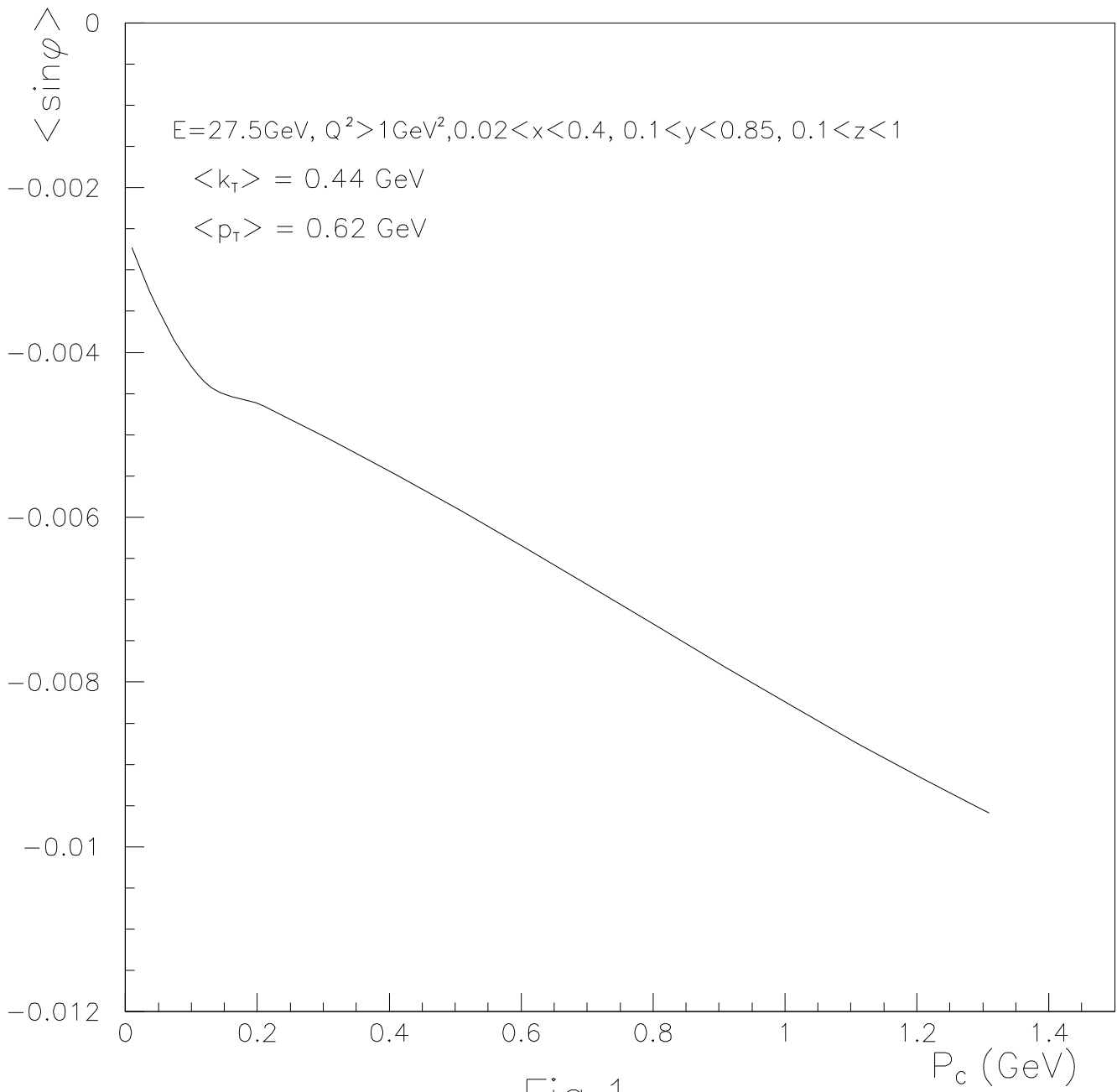
\end{document}